\RequirePackage{fix-cm}
\documentclass[a4paper,11pt]{article}

\makeatletter

\newcommand{\Rmnum}[1]{\expandafter\@slowromancap\romannumeral #1@}
\makeatother

\newlength{\Oldarrayrulewidth}

\pdfoutput=1 

\usepackage{jheppub} 
\usepackage{multirow}
\usepackage[T1]{fontenc} 
\usepackage{graphicx}
\usepackage{placeins}

\DeclareMathSizes{10.95}{11.3}{6}{4.5}

\title{\boldmath Di-Higgs Production in SUSY models at the LHC}

\author{Peisi Huang, and Yu Hang Ng}

\affiliation{Department of Physics and Astronomy, University of Nebraska-Lincoln, Lincoln, NE, 68588}

\emailAdd{peisi.huang@unl.edu, yu-hang.ng@huskers.unl.edu}

\abstract{We study the modification to di-Higgs production via gluon fusion within the context of the Minimal Supersymmetric Standard Model(MSSM) and the Next-to-Minimal Supersymmetric Standard Model(NMSSM) in the parameter space allowed by current experimental and theoretical constraints, and also relevant to the Large Hadron Collider(LHC) experiments in the near future. The calculation is based on the analytical expression of the leading order Feynman amplitudes (which includes both quark and squark loops). We separate the di-Higgs production cross section into resonant, non-resonant, and interference parts, in order to better understand the mechanisms that are responsible for the modification to di-Higgs production rate in different regions of the allowed parameter space. We also investigate the sensitivity of High-Luminosity LHC (HL-LHC) to the di-Higgs production in these low energy supersymmetry(SUSY) models. Furthermore, we examine the complementarity between di-Higgs searches and direct searches for BSM particles and precision Higgs couplings measurements at the HL-LHC. We found that the di-Higgs production cross section can be enhanced significantly through resonant production. In the region where the resonant production cross section is small, di-Higgs receives a moderate enhancement due to the modifications in the Higgs couplings. In addition, there is a strong correlation between di-Higgs production and single Higgs production, and di-Higgs is not as sensitive as single Higgs at the HL-LHC. }

\begin{document} 
\maketitle
\flushbottom

\section{Introduction}
\label{sec:intro}
The Higgs boson discovery by the CMS and ATLAS experiments in 2012~\cite{higgs2012Atlas,higgs2012CMS} is the first and a crucial step in understanding the mechanism of the electroweak symmetry breaking. The Higgs properties, including the mass, spin, parity, and its couplings to other Standard Model (SM) particles have been measured with precisions in the subsequent analysis~\cite{higgsSPCMS,higgsSPAtlas,HcouplingsCMS,HcouplingsAtlas}. However, we know very little about the Higgs potential. In the SM, the Higgs potential is
\begin{equation}
    V = -\mu^2 \phi^{\dagger}\phi + \lambda (\phi^{\dagger}\phi)^2,
\end{equation}
which is completely specified by two parameters $\mu$ and $\lambda$. $\mu$ and $\lambda$ can be determined from the vacuum expectation value (vev) of the Higgs field, and the mass of the Higgs boson, but there is no direct measurement beyond that.   
Thus, the next step in understanding the shape of the Higgs potential is to measure the Higgs trilinear coupling. The Higgs trilinear coupling can be probed by the di-Higgs production at the LHC.  
\par
Di-Higgs production is sensitive to new physics, including new scalar resonances~\cite{Chen:2014ask,Huang:2015tdv,Lewis:2017dme,DiMicco:2019ngk}, new colored particles~\cite{Batell:2015koa,Dawson:2015oha,Huang:2017nnw,DiMicco:2019ngk}, and modified Higgs couplings~\cite{Chen:2017qcz,Huang:2015tdv,Huang:2017nnw,DiMicco:2019ngk}, and therefore can be complementary to direct searches of new particles, and precision Higgs coupling measurements. In particular, in this paper, we consider di-Higgs production with low energy supersymmetry (SUSY) models. 

In low energy SUSY models, the di-Higgs production rate can be modified through various mechanisms. First, the Higgs sector is extended by introducing an additional Higgs doublet, in the Minimal Supersymmetric extension of the SM (MSSM) and an additional singlet in the Next-to-Minimal Supersymmetric extension of the SM (NMSSM). The additional neutral Higgs states can be produced at the LHC and can decay into a pair of SM-like Higgs bosons, and therefore contribute to di-Higgs production. Second, the low energy SUSY models allow the presence of new light colored particles coupled strongly to the Higgs. Those new colored particles give new QCD loop diagrams contributing to di-Higgs production~\cite{,Batell:2015koa,Huang:2017nnw}. There will be new interference terms arise from those new diagrams as well~\cite{Basler:2019nas,Carena:2018vpt}. Third, in SUSY models, Higgs couplings, including Higgs couplings to SM particles, and the Higgs self-couplings, may present small deviations with respect to the SM ones, resulting in modified di-Higgs production~\cite{Huang:2017nnw}. The modified couplings also change the decay of the Higgs, resulting in additional modifications in specific channels. In this paper, we study the di-Higgs production in the MSSM and the NMSSM, with a focus on identifying the dominant contribution in different regions of parameter space. We also study the complementarity between di-Higgs searches and other searches, such as direct searches for new scalars, and precision Higgs measurements at the HL-LHC.   
\par
This article is structured as follows. In section 2, we calculate the leading order di-Higgs production cross section in the MSSM and the NMSSM.
In section 3, we present the main results of this paper. First, we identify the parameter space that satisfies current experimental constraints and theoretical requirements. Then, we show the results of di-Higgs production cross section in the $b\bar{b}$$\tau^{+}\tau^{-}$ and $b\bar{b}$$\gamma\gamma$ final states, in those regions. Furthermore, we study the complementarity of di-Higgs with other searches. We also discuss the resonant, non-resonant, and interference contributions of the di-Higgs production in different regions of the parameter space. Finally, we reserve section 4 for a summary of the main results.   

\begin{figure}[!htb]
\centering
\includegraphics[width=\textwidth]{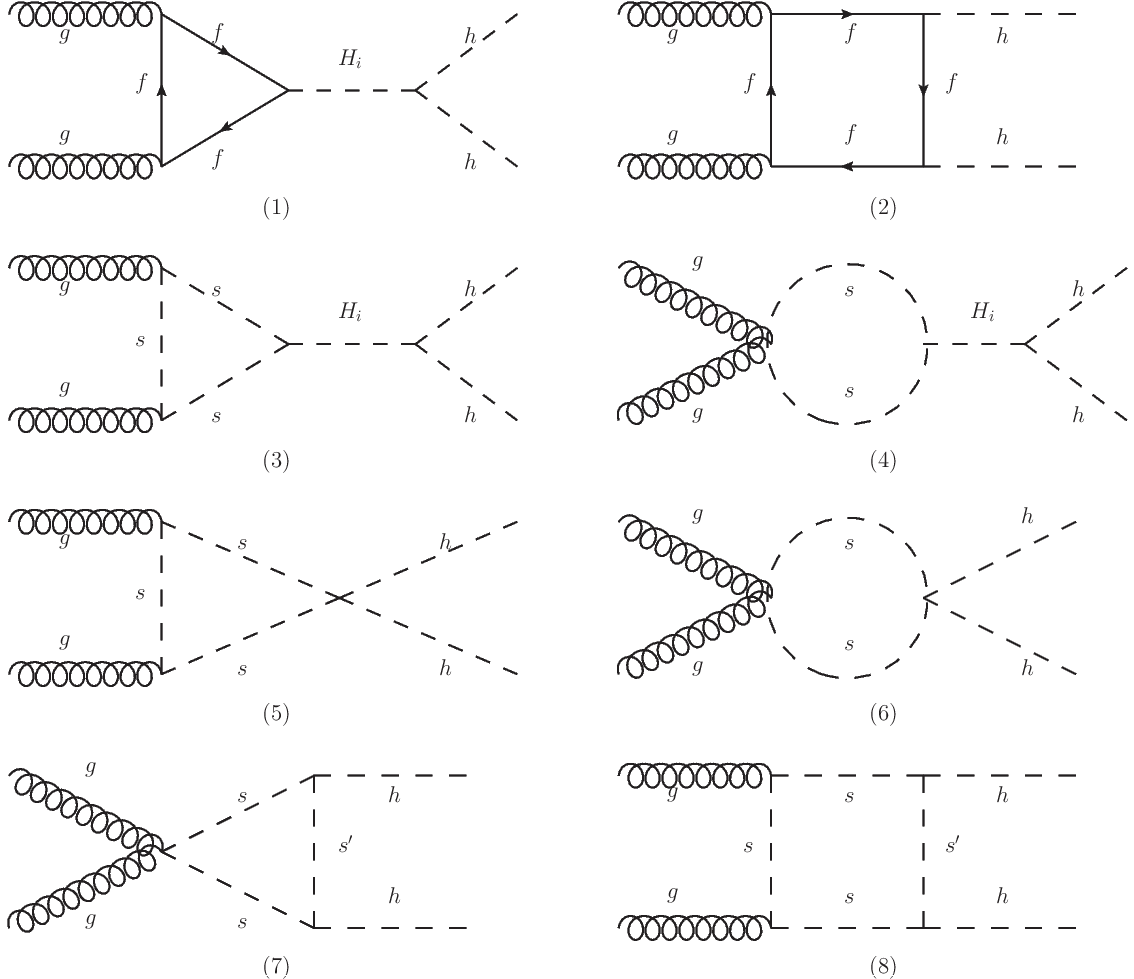}
\caption{\label{fig:1} Leading order Feynman diagrams for di-Higgs production in the MSSM at LHC. $H_i$ are CP even Higgs bosons, $f$ are top and bottom quarks, and $s,s'$ are stops.}
\end{figure}


\section{Matrix Elements and Cross Sections}
The dominant di-Higgs production mode at the LHC is the gluon-gluon fusion (ggF) process. In the SM, there are two diagrams contributing to the di-Higgs production at the leading order, the triangle diagram (diagram (1) in Fig.~\ref{fig:1}, with $H_i = h$), and the box diagram (diagram (2) in Fig.~\ref{fig:1}). In this section, we calculate the leading order di-Higgs production cross section $ \sigma (pp \rightarrow gg \rightarrow hh)$ at $\sqrt{s}=14$ TeV by using the analytical expressions for one-loop amplitudes of $gg \rightarrow hh$ in the MSSM and the NMSSM.
\subsection{MSSM}
The leading order Feynman diagrams contributing to the $gg \rightarrow hh$ process in the MSSM at the LHC are summarized in Fig.~\ref{fig:1}. In the MSSM, there are two neutral CP-even Higgs bosons and one neutral CP-odd Higgs boson. In this work, we assume that the lighter CP-even neutral Higgs boson to be SM-like, and denote the SM-like Higgs boson as $h$, and the heavy CP-even neutral heavy Higgs boson as $H$. The heavy Higgs can be produced through quark and squark loops and can decay to a pair of SM Higgs bosons, as shown in diagrams (1), (3), and (4) in Fig.~\ref{fig:1} with $H_i = H$. The squarks also lead to new diagrams that contribute to di-Higgs production, as shown in diagrams (3)-(8) in Fig.~\ref{fig:1}.  
The corresponding spin and color averaged partonic differential cross section is
\begin{equation}\label{eq:2}
\dfrac{d\hat{\sigma}}{d\hat{t}}=\dfrac{(\alpha_{s}(Q))^{2}}{2^{13}\pi^{3}}(|A|^{2}+|B|^{2}) \,,
\end{equation}
where the matrix elements are separated by the helicities of the initial gluons. $A_{(i)}$ are matrix elements of the $i$-th diagram in Fig.\ref{fig:1} when helicities of initial gluons are identical, and $B_{(i)}$ are matrix elements of the $i$-th diagram in Fig.\ref{fig:1} when helicities of initial gluons are opposite. A and B denote the sum of those matrix elements. 
\begingroup
\allowdisplaybreaks
\begin{align}
A &= \sum_{H_{i}=h,H} \sum_{f=b,t} A_{(1)}(f,H_{i}) + \sum_{f=b,t} A_{(2)}(f) + \sum_{H_{i}=h,H} \sum_{s=\tilde{t}_{1},\tilde{t}_{2}} A_{(3+4)}(s,H_{i})          \nonumber
\\ 
  &+ \sum_{s=\tilde{t}_{1},\tilde{t}_{2}} A_{(5+6)}(s) + \sum_{s^{'}=\tilde{t}_{1},\tilde{t}_{2}} \sum_{s=\tilde{t}_{1},\tilde{t}_{2}} A_{(7+8)}(s,s') \,, \nonumber
\\
B &= \sum_{f=b,t} B_{(2)}(f) + \sum_{s^{'}=\tilde{t}_{1},\tilde{t}_{2}} \sum_{s=\tilde{t}_{1},\tilde{t}_{2}} B_{(8)}(s,s') \,, \nonumber
\end{align}
\endgroup
The matrix elements $A_{(i)}$ and $B_{(i)}$ read,
\begingroup
\allowdisplaybreaks
\begin{align}
A_{(1)}(f,H_{i}) &= \dfrac{g_{H_{i}hh}g_{H_{i}ff}}{m_{f}}\dfrac{F_{\triangleright}^{(1/2)}(m_{f})}{\hat{s}-m_{H_{i}}^{2}+i\Gamma_{H_{i}}m_{H_{i}}} \label{eq:3.1}
\\
A_{(2)}(f) &= \left( \dfrac{g_{hff}}{m_{f}} \right)^{2} F_{\square}^{(1/2)}(m_{f}) \label{eq:3.2}
\\
B_{(2)}(f) &= \left( \dfrac{g_{hff}}{m_{f}} \right)^{2} G_{\square}^{(1/2)}(m_{f}) \label{eq:3.3}
\\
A_{(3+4)}(s,H_{i}) &= \dfrac{g_{H_{i}hh}g_{H_{i}ss}}{2m_{s}^{2}}\dfrac{F_{\triangleright}^{(0)}(m_{s})}{\hat{s}-m_{H_{i}}^{2}+i\Gamma_{H_{i}}m_{H_{i}}} \label{eq:3.4}
\\
A_{(5+6)}(s) &= -\dfrac{g_{hhss}}{2m_{s}^{2}} F_{\triangleright}^{(0)}(m_{s}) \label{eq:3.5}
\\
A_{(7+8)}(s,s') &= \dfrac{g_{hss'}^{2}}{2 m_{s}^{2} m_{s'}^{2}} F_{\square}^{(0)}(m_{s},m_{s'}) \label{eq:3.6}
\\
B_{(8)}(s,s') &= \dfrac{g_{hss'}^{2}}{2 m_{s}^{2} m_{s'}^{2}} G_{\square}^{(0)}(m_{s},m_{s'}) \label{eq:3.7}
\end{align}
\endgroup
 Expressions for the form factors ($F_{\triangleright}$, $F_{\square}$, $G_{\square}$) are given in Appendix A. $\hat{s}$, $\hat{t}$, $\hat{u}$ are Mandelstam variables. We include the dominant fermion contributions from top and bottom loops, and sfermion contributions from stops and sbottoms. The MSSM couplings($g$), masses($m_{H_{i}}$), and decay widths($\Gamma_{H_{i}}$) of Higgs bosons are calculated by using \textbf{FeynHiggs-2.14.3} ~\cite{FH98_1,FH98_2,FH02,FH06,FH13,FH16,FH17}. We choose $Q=m_{hh}/2$, where $m_{hh}$ is the invariant mass of the SM-like Higgs pair.
\par
We separate the differential cross sections into resonant, non-resonant, and interference contributions as
\begin{align}
\dfrac{d\hat{\sigma}_{res}}{d\hat{t}}&=\dfrac{(\alpha_{s}(Q))^{2}}{2^{13}\pi^{3}}|A_{res}|^{2} \,, \label{eq:4.1}
\\
\dfrac{d\hat{\sigma}_{nr}}{d\hat{t}}&=\dfrac{(\alpha_{s}(Q))^{2}}{2^{13}\pi^{3}}(|A_{nr}|^{2}+|B_{nr}|^{2}) \,, \label{eq:4.2}
\\
\dfrac{d\hat{\sigma}_{int}}{d\hat{t}}&=\dfrac{(\alpha_{s}(Q))^{2}}{2^{13}\pi^{3}}(2\operatorname{Re}(A_{res} \cdot A_{nr}^{\scalebox{.7}{*}})) \,, \label{eq:4.3}
\end{align}
where
\begin{align}
A_{res} &= \sum_{f=b,t} A_{(1)}(f,H) + \sum_{s=\tilde{t}_{1},\tilde{t}_{2}} A_{(3+4)}(s,H) \,, \nonumber
\\
A_{nr} &= A-A_{res} \,, \nonumber
\\
B_{nr} &= B \,. \nonumber
\end{align}
The resonant amplitude is defined such that it has a pole in the region of interest. The non-resonant amplitude includes all other Feynman diagrams, and the interference cross section corresponds to the interference between resonant and non-resonant amplitude, as the name suggests.
\par
The total cross section corresponds to each partonic cross section can be obtained by using
\begin{equation}\label{eq:5}
\sigma(pp \rightarrow gg \rightarrow hh) = \int_{(2m_{h})^{2}}^{(14 \mathrm{TeV})^{2}} d\hat{s} \dfrac{d \mathcal{L}_{gg}}{d \hat{s}} \hat{\sigma} \,,
\end{equation}
where $\dfrac{d \mathcal{L}_{gg}}{d \hat{s}}$ is the differential gluon-gluon luminosity as defined in~\cite{Campbell2006wx}.

\subsection{NMSSM}
The NMSSM has one extra supermultiplet as compared to the MSSM, i.e. the gauge singlet chiral superfield $\hat{S}$, which contains one complex spin-0 singlet ($S$) and one spin-1/2 singlino ($\tilde{S}$). $S$ gives rise to one neutral CP-even Higgs and one neutral CP-odd Higgs. In this article, we only consider the $\mathbb{Z}_{3}$-invariant NMSSM, which has the scale invariant superpotential $W_{\mathrm{Higgs}}$ given by
\begin{equation}\label{eq:6}
W_{\mathrm{Higgs}} = \lambda \hat{S} \hat{H}_{u} \cdot \hat{H}_{d} + \dfrac{\kappa}{3} \hat{S}^{\,3} \,,
\end{equation}
where $W_{\mathrm{Higgs}}$ is the part of superpotential that depends exclusively on Higgs superfields $\hat{H}_{u}$, $\hat{H}_{d}$, and $\hat{S}$. In the next paragraph, we will briefly introduce the notations and sign conventions of relevant NMSSM parameters that appear in this paper. Please refer to the review paper~\cite{Ellwanger2009dp} for more details about the NMSSM.   
\par
The first term in Eq.~(\ref{eq:6}) replaces the $\mu \hat{H}_{u} \cdot \hat{H}_{d} $ term in MSSM superpotential, and generates the effective $\mu$-term at electroweak scale when $S$ acquires a non-vanishing vev:
\begin{equation}\label{eq:7.1}
\mu_{\mathrm{eff}} = \lambda \langle S \rangle \,.
\end{equation}
$v_{u}$ and $v_{d}$ are vevs of $H_{u}^{0}$ and $H_{d}^{0}$ respectively, and 
\begin{equation}\label{eq:7.2}
\tan \beta = \dfrac{v_{u}}{v_{d}} \,.
\end{equation}
$A_{\lambda}$ and $A_{\kappa}$ are trilinear soft SUSY breaking couplings correspond to the first and second term in eq.(\ref{eq:6}) respectively. $M_{1}$, $M_{2}$, and $M_{3}$ are soft SUSY breaking masses of $U(1)_{Y}$ gaugino, $SU(2)$ gauginos, and $SU(3)$ gauginos respectively. We choose $\tan \beta$ and $\lambda$ to be positive, while $\mu_{\mathrm{eff}}$, $\kappa$, $A_{\lambda}$, $A_{\kappa}$ can have both signs.

\par
The set of leading order Feynman diagrams for $gg \rightarrow hh$ in NMSSM is the same as the MSSM case, except that there are three neutral CP-even Higgs bosons ($H_{i} = H_{1},H_{2},H_{3}$, ordered by ascending mass) in NMSSM. Thus, Eq.~(\ref{eq:3.1}-\ref{eq:3.7}) are still valid in NMSSM case. The NMSSM couplings, masses, and decay rates of Higgs bosons are calculated by using \textbf{NMSSMTools-5.4.1} ~\cite{NT04,NT05,NT09,NT97,NT07,NT05DM}. 

\section{Phenomenological Study}

\subsection{MSSM}

\begin{figure}[tb]
\centering
\includegraphics[scale=0.3]{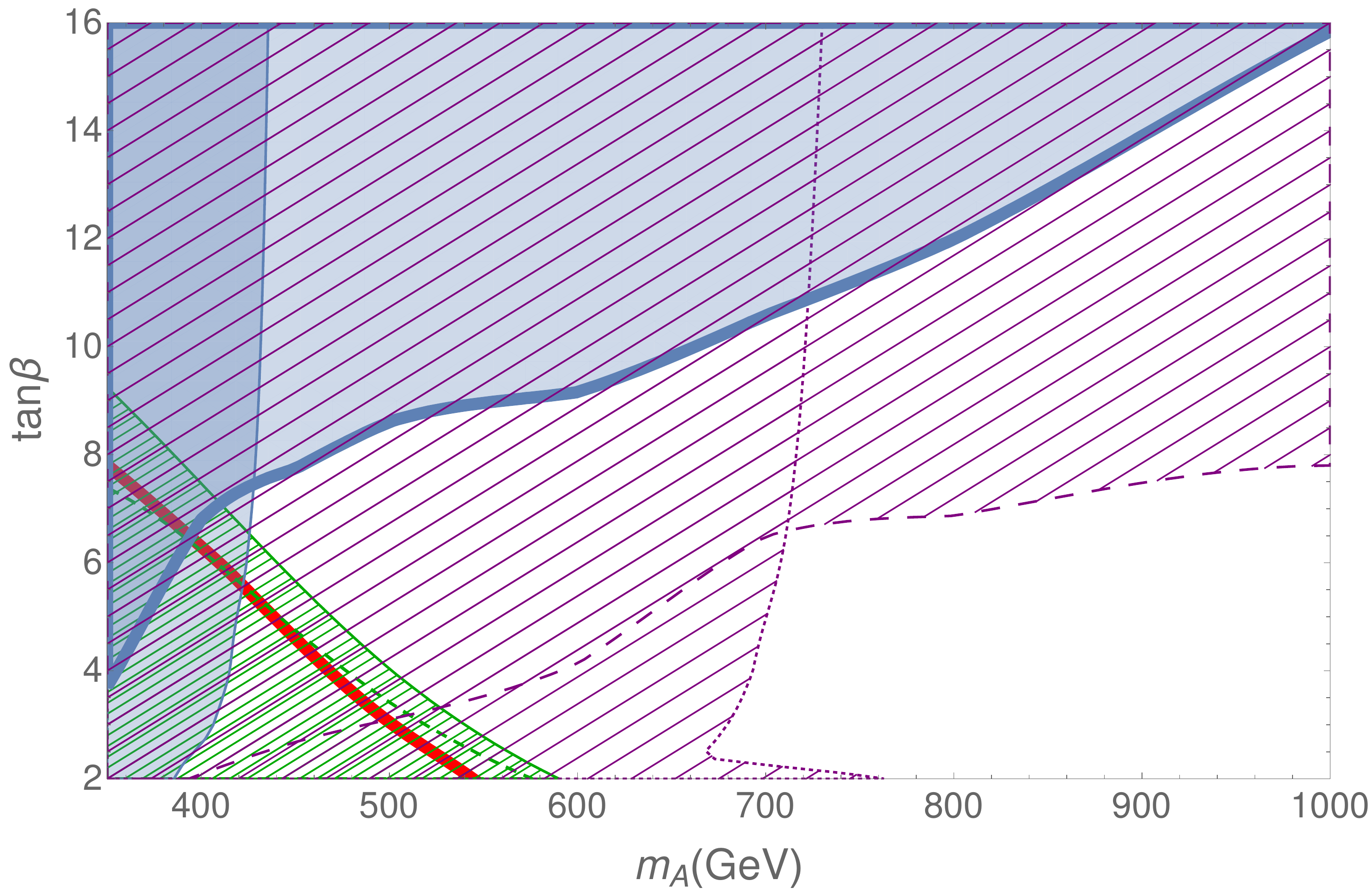}
\caption{\label{fig:2} The MSSM parameter space with current experimental constraints and projected sensitivities. The blue shaded region is excluded by current experiment constraints. The purple and green shaded region can be reached by HL-LHC. The red solid line represents the case when resonant cross section equals to non-resonant cross section.}
\end{figure}

In the MSSM study, for the stops to contribute in a relevant way, the stop should not be too far away from the weak scale~\cite{Huang:2017nnw}. Therefore, we chose the lighter stop mass to be 600~GeV, which is right above the current LHC limit ~\cite{mstAtlas,mstCMS}. To allow for large mixings in the stop sector, which can enhance the di-Higgs production rate through diagrams (7) and (8) in Fig.~\ref{fig:1}, we choose the heavier stop mass to be 5 ~TeV. The stop mixing parameter $X_t$ is chosen so that the SM-like Higgs mass is 125 GeV. 
We calculated the SM-like Higgs mass using \textbf{FeynHiggs-2.14.3} ~\cite{FH98_1,FH98_2,FH02,FH06,FH13,FH16,FH17}, and found that in the region of interest, there always exists a value of $X_{t}$ such that $m_{h}=125$ GeV. 
	 Futhermore, $X_{t}$ cannot be too large in order to have a stable SM-like vacuum. Here, we use the approximate bound~\cite{vacstabbound}.
\begin{equation}\label{eq:9}
(X_{t}+\dfrac{\mu}{\tan\beta})^{2} \leq (3.4(m^{2}_{\tilde{Q}_{3}}+m^{2}_{\tilde{t}_{R}})+0.5 |m^{2}_{\tilde{Q}_{3}}-m^{2}_{\tilde{t}_{R}}|)+60(\dfrac{m^{2}_{Z}}{2}\cos2\beta+m^{2}_{A}\cos^{2}\beta) \,.
\end{equation}
We checked that this bound is satisfied by every point in the parameter space that is shown in Fig.\ref{fig:2}. 

\par Then we fix the masses and the mixings in the sbottom sector by imposing $m_{\tilde{t}_{R}} = m_{\tilde{b}_{R}}$, and $A_{t} = A_{b}$.
We ignore the contributions from all other sfermions and electroweakinos. When the electroweakinos are lighter than $m_{H}/2$, the heavy Higgs starts to decay to a pair of electroweakinos, which reduces its branching ratio to a pair of SM Higgs, and therefore reduces the resonant di-Higgs production cross section. As we aim to understand how large the di-Higgs production cross section can be at the LHC, we ignore the contributions from electroweakinos. 
\par
Based on those considerations, we keep $m_{A}$, the mass of the CP-odd Higgs boson,  and $\tan \beta$, the ratio between two vevs, as free parameters in our study.
The ranges of $m_{A}$ and $\tan \beta$ are restricted by several experimental observations and theoretical requirements as shown in Fig.\ref{fig:2}. First, the Higgs couplings depend strongly on the mixing between the SM-like Higgs and the new CP-even Higgs, and in the region of interest, can deviate significantly from the SM. Therefore, the parameter space is restricted by precision Higgs measurements, and the most stringent limit comes from the Higgs boson coupling to bottom quark measurement at ATLAS~\cite{HcouplingsAtlas}. The blue shaded region with the thin solid boundary line is excluded by the precision measurement of Higgs couplings at $\sqrt{s}$ = 13 TeV ~\cite{HcouplingsAtlas,HcouplingsCMS}. Second, the heavy Higgs couples to the SM particles, and can be searched for through its decays. The most sensitive channel is in the di-tau final state, and the blue shaded region with the thick solid boundary line is excluded by the search for the additional neutral Higgs bosons in the di-tau final state at $\sqrt{s}$ = 13 TeV ~\cite{HHditauCMS,HHditauAtlas}. 

\par
\begin{figure}[tb]
\centering
\includegraphics[width=0.475\textwidth, height=0.2\textheight]{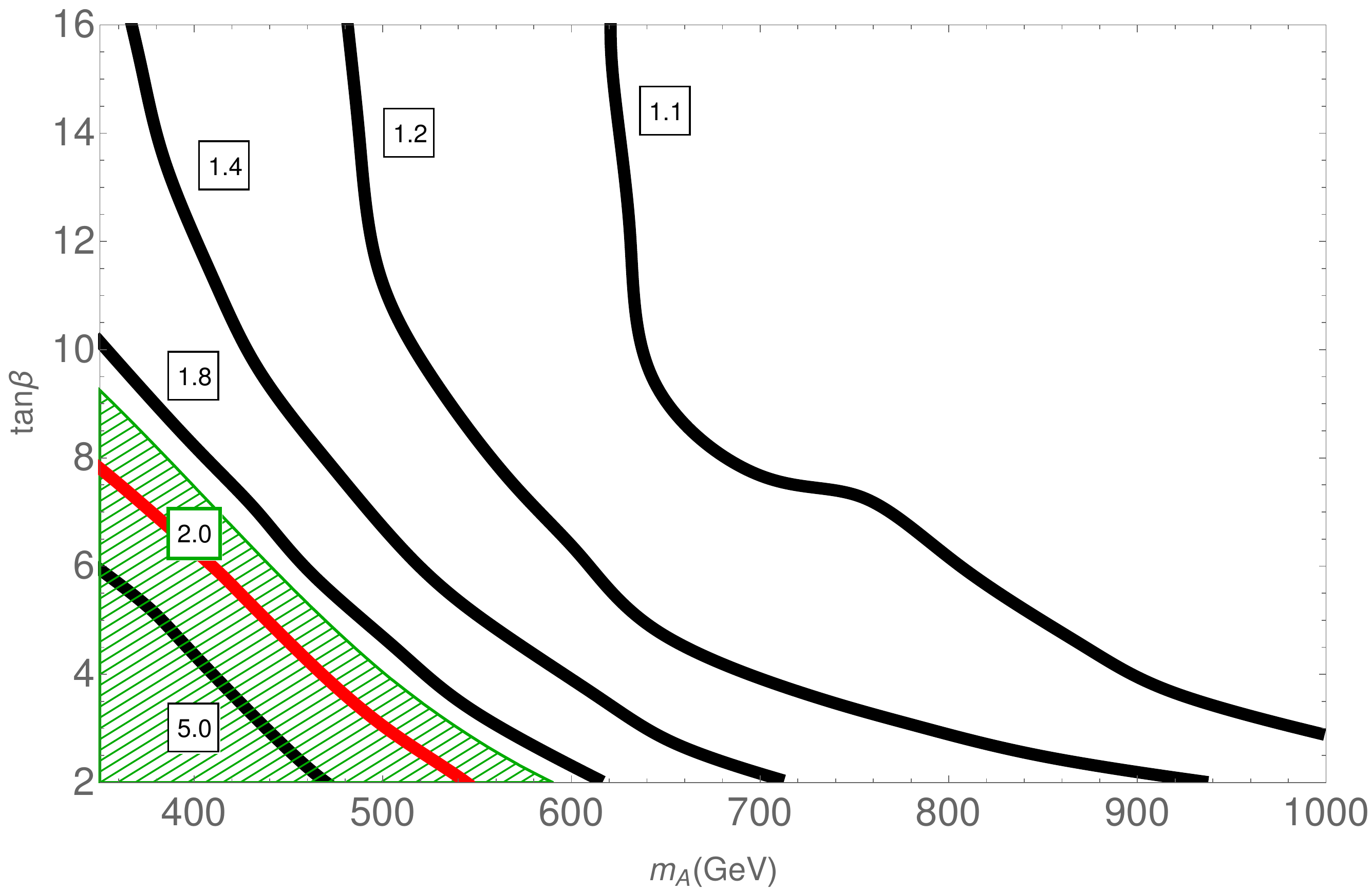}
\hfill
\includegraphics[width=0.475\textwidth, height=0.2\textheight]{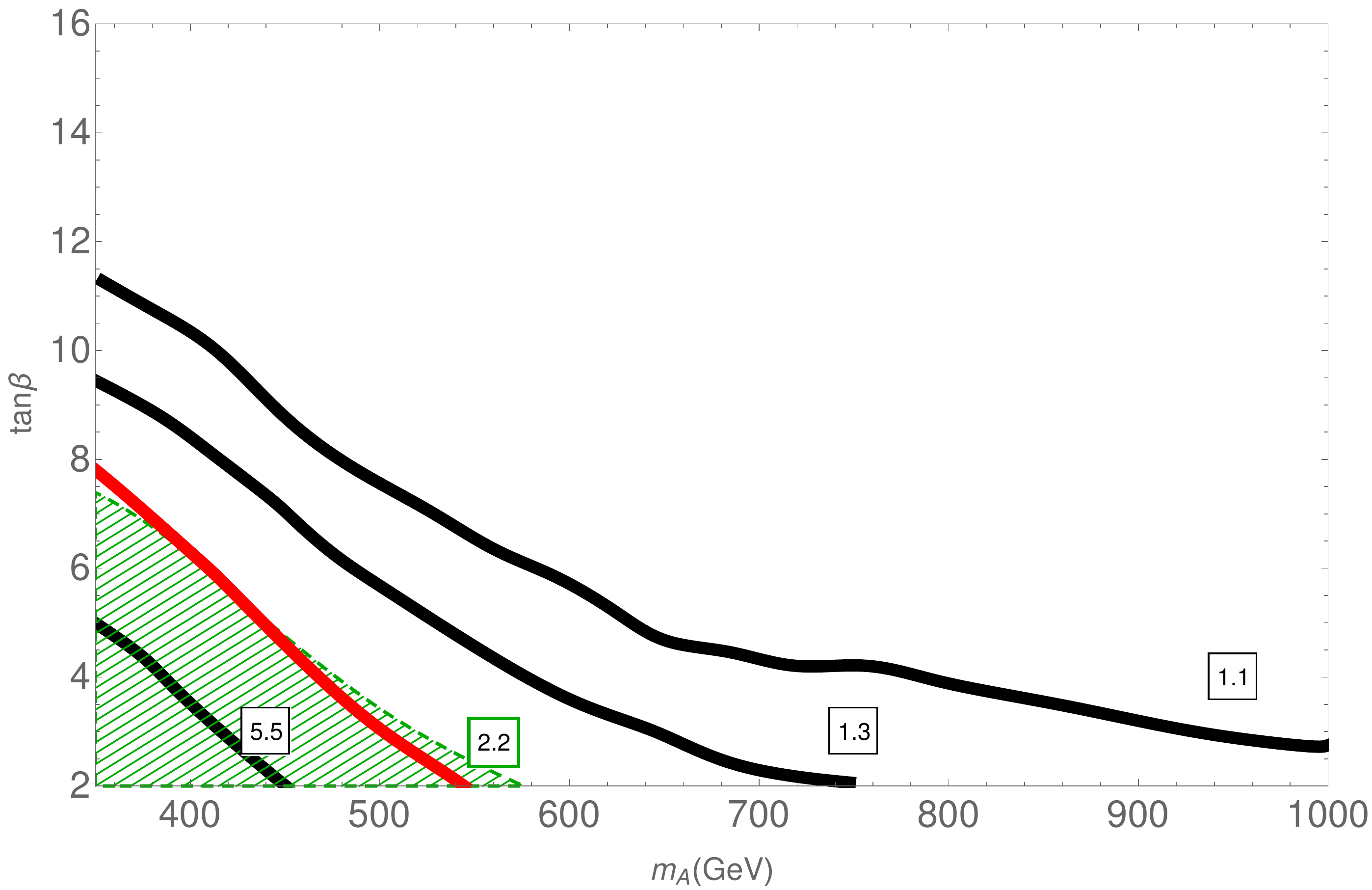}
\caption{\label{fig:3} Contour plots for $\sigma(pp \rightarrow hh \rightarrow b\bar{b}\tau^{+}\tau^{-})$ (left panel) and $\sigma(pp \rightarrow hh \rightarrow b\bar{b}\gamma\gamma)$ (right panel) normalized to SM value at $\sqrt{s}$ = 14 TeV. The cross sections in the shaded regions are outside the 95\% CL of the projected di-Higgs production cross section at HL-LHC.}
\end{figure}
\begin{figure}[tb]
\centering
\includegraphics[scale=0.4]{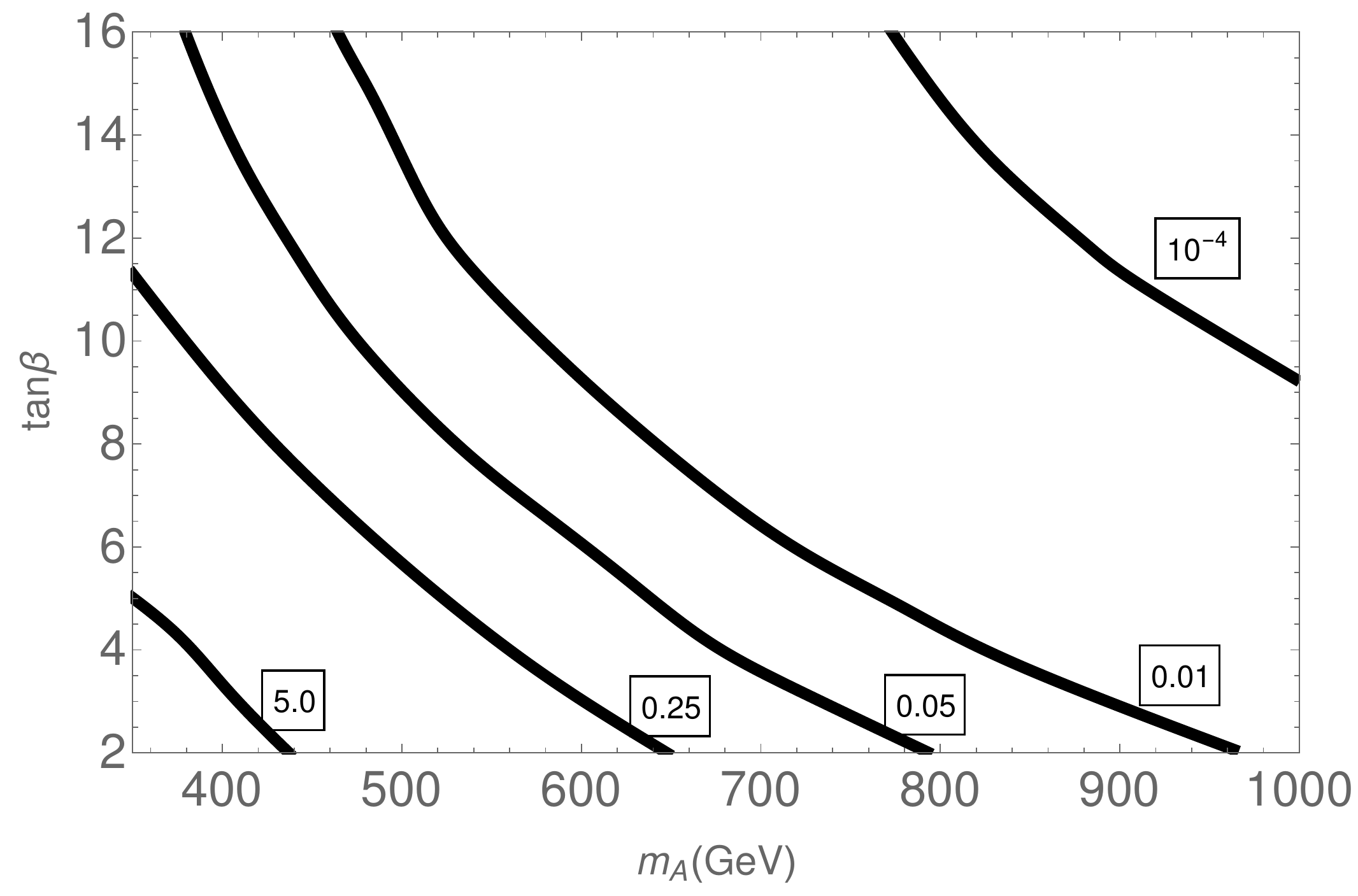}
\caption{\label{fig:4} Resonant contribution of di-Higgs production cross section (normalized to SM value) at $\sqrt{s}$ = 14 TeV.}
\end{figure}

We calculate the leading order di-Higgs production cross section as described in the previous section, and the results are shown in Fig.~\ref{fig:3}. 
To account for the modifications in the Higgs decay, we show the results in the two most sensitive channels, $bb\gamma\gamma$ and $bb\tau\tau$, according to the ATLAS projections~\cite{HHditauAtlasHL}. In the region of interest, $B(h \rightarrow b\bar{b})B(h \rightarrow \tau^{+}\tau^{-})$ is always enhanced, and it can be enhanced up to 25\% compared to the SM value, while $B(h \rightarrow b\bar{b})B(h \rightarrow \gamma\gamma)$ is suppressed in most of the parameter space, and it can be suppressed up to 14\% compared to the SM value. The di-Higgs production rate can be enhanced significantly in the region where $m_{A}$ and $\tan \beta$ are small. For instance, when $m_{A}=400$ GeV and $\tan \beta=2$, $\sigma(pp \rightarrow hh \rightarrow b\bar{b}\tau^{+}\tau^{-})/\mathrm{SM} = 13.7$ and $\sigma(pp \rightarrow hh \rightarrow b\bar{b}\gamma\gamma)/\mathrm{SM} = 11.3$. This enhancement is mainly due to the large resonant contribution of di-Higgs production in this region, which can also be seen in Fig.\ref{fig:4}. 
In Fig.~\ref{fig:4}, we plot the resonant production of di-Higgs normalized to the SM production rate. The production rate can be a few of the SM value when $m_A$ and $\tan\beta$ are small, and decreases quickly as $m_A$ and $\tan\beta$ increase as expected.

\par
\begin{figure}[tb]
\centering
\includegraphics[scale=0.33]{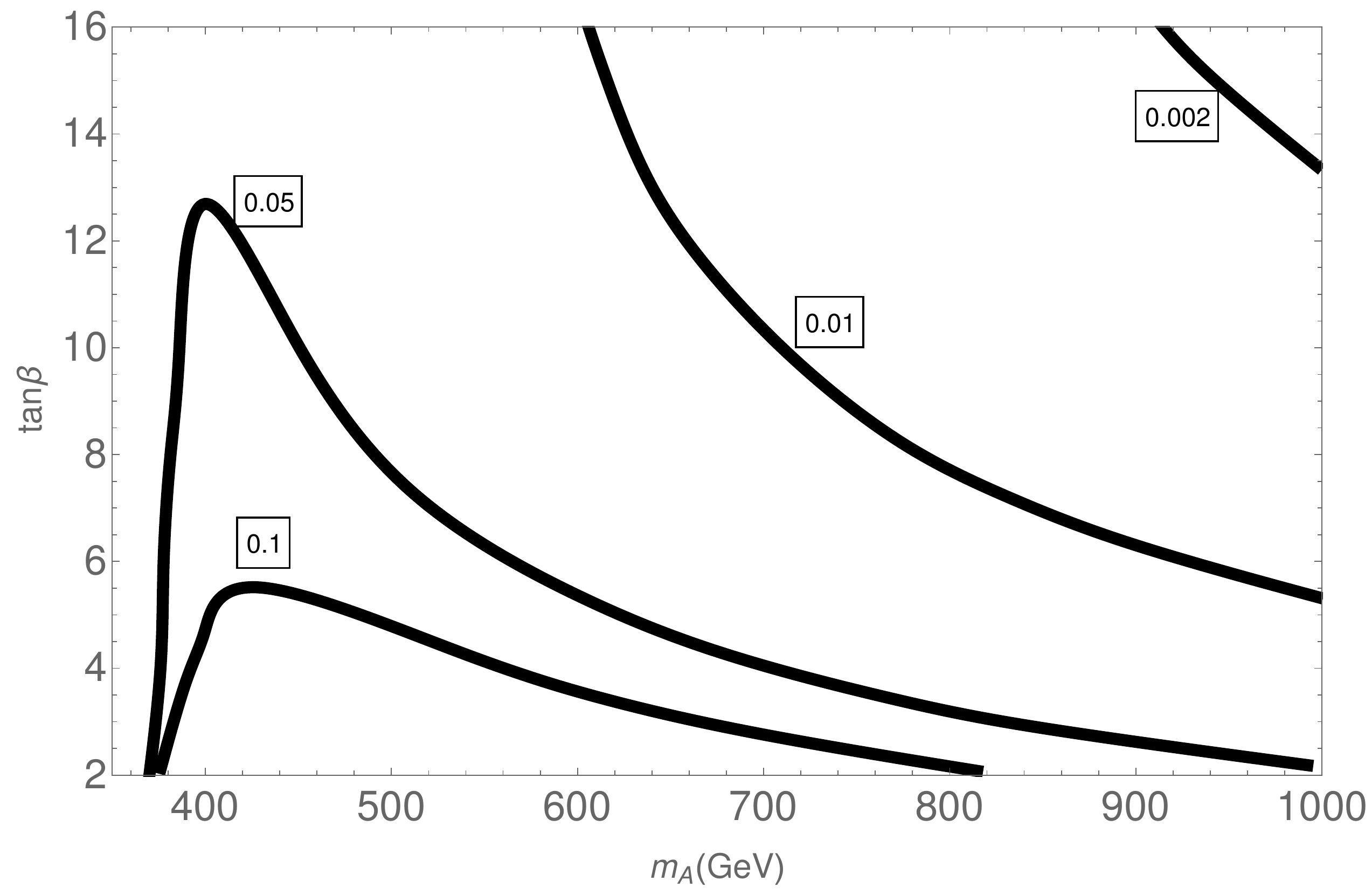}
\caption{\label{fig:5} Interference (normalized to SM value) between resonant and non-resonant amplitude at $\sqrt{s}$ = 14 TeV.}
\end{figure}

 Outside the region where di-Higgs production receives a large correction from the resonant production, the modifications come from the non-resonant production. The non-resonant di-Higgs production is very close to the SM contribution in the whole region of interest. Our calculations show that $\kappa_{t} = g_{htt}/g_{htt}^{\mathrm{SM}}$ varies between 0.95 and 1, while $\lambda_{3} = (g_{hhh}-g_{hhh}^{\mathrm{SM}})/g_{hhh}^{\mathrm{SM}}$ varies between -0.24 and -0.12 in the parameter space of interest. The decrease in the SM-like Higgs self-coupling reduces the destructive interference between the triangle diagram and the fermionic box diagram (i.e. first and second diagram in Fig.\ref{fig:1}), hence it enhances non-resonant di-Higgs production cross section. However, this effect is offset by a small decrease in $\kappa_{t}$. The box diagram, which dominates over the triangle diagram, is proportional to $\kappa_t^4$, so the di-Higgs rate is very sensitive to the value of $\kappa_t$. Therefore, a small decrease in $\kappa_t$ offsets the decrease in $\lambda_3$. As a result, the non-resonant di-Higgs cross section is only larger than the SM di-Higgs production cross section by a few percent in most of the parameter space. The interference between the resonant and non-resonant amplitude is small over the whole parameter space, which we plot in Fig.\ref{fig:5}.
Contribution from stop loops is less than 3\% of the total di-Higgs production cross section.
Therefore, in the MSSM case, the major modification to di-Higgs production is from the new resonance, which is the new CP-even neutral Higgs, and the modifications to the branching ratios.
\par
We also study the complementarity between di-Higgs and other Higgs studies. In Fig.~\ref{fig:2}, we compare the projected sensitivity in the $bb\gamma\gamma$ and the $bb\tau\tau$ channels with other searches. We separate the parameter space into two regions, the one below the red solid line is dominated by resonant production, while the region above the red solid line is dominated by non-resonant production. The green shaded region with solid boundary line and the dashed boundary line show the region that can be excluded by the non-resonant $bb\tau\tau$ and $bb\gamma\gamma$ searches at HL-LHC with 3000fb$^{-1}$ respectively. The projected sensitivity for the resonant production is not available at present, but we do expect that is better compared to the sensitivity for the non-resonant production. Therefore, the whole green shaded region can be probed by HL-LHC.  
The purple shaded region with the dotted boundary line shows the region can be tested by the projected Higgs boson couplings measurements at HL-LHC ~\cite{HcouplingsAtlasHL}.
The purple shaded region with the dashed boundary line shows the projected reach for the additional neutral Higgs bosons in the di-tau final state at HL-LHC at 95$\%$ C.L. ~\cite{HHditauAtlasHL}.  From Fig.\ref{fig:2}, a large fraction of the parameter space that is not excluded by current experimental constraints will be tested by the HL-LHC experiments (i.e. the purple and green region). The di-Higgs production is not as sensitive as other searches, but provides a complementary probe at the HL-LHC.

\subsection{NMSSM}
\begin{figure}[tb]
\centering
\includegraphics[scale=0.4]{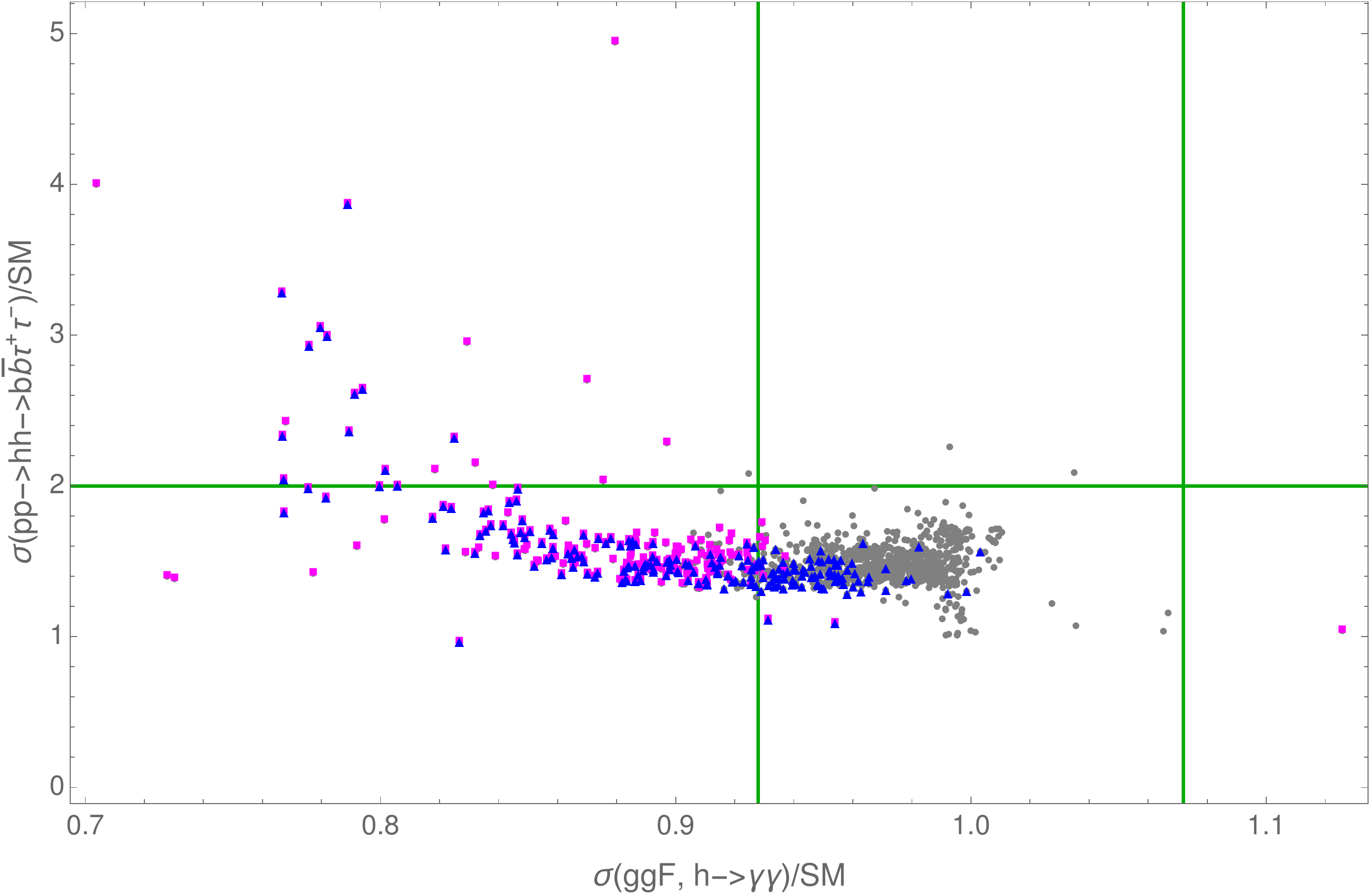}

\vspace{.5 cm}

\includegraphics[scale=0.44]{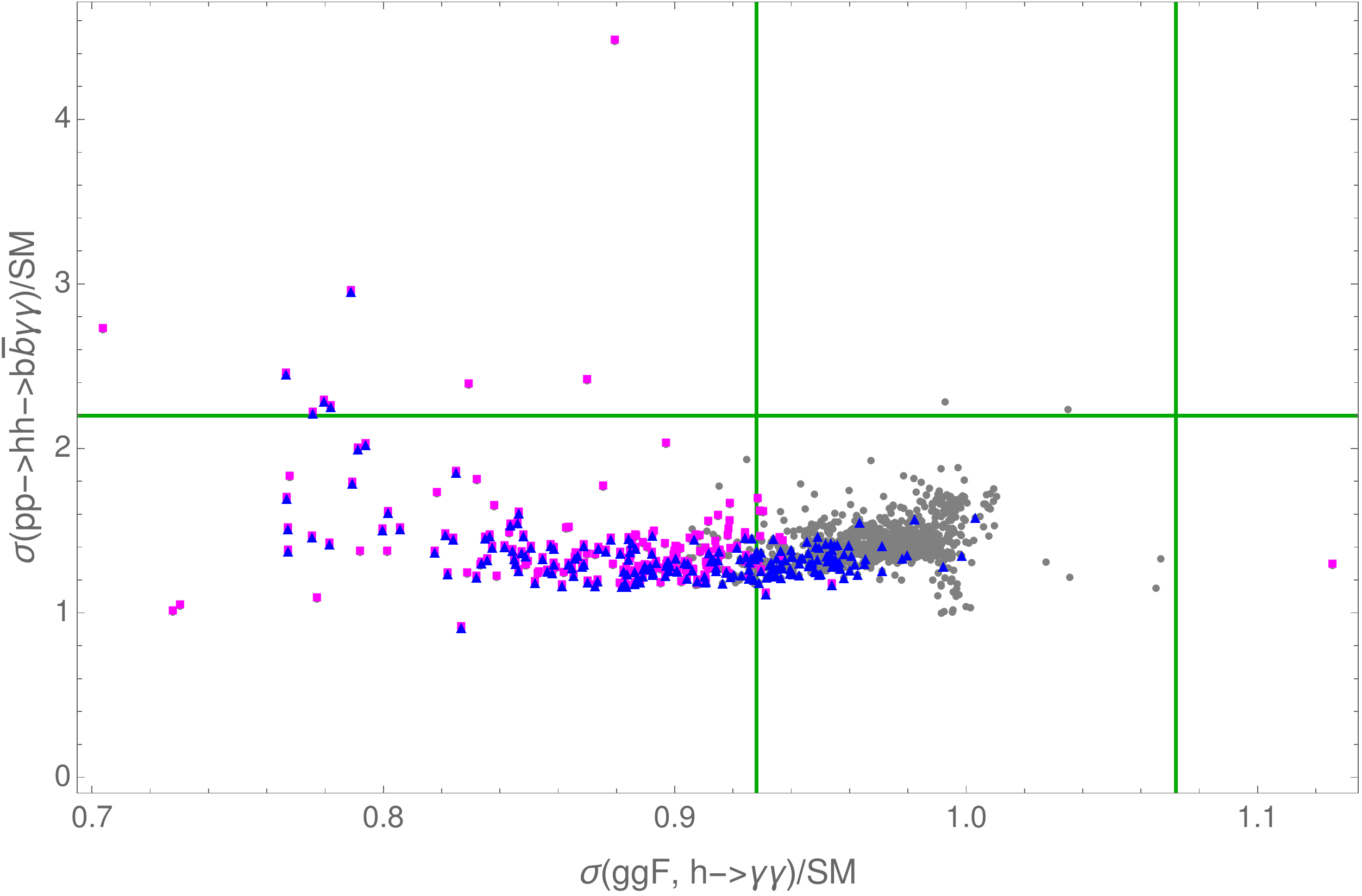}
\caption{\label{fig:6} Scatter plots for $\sigma(pp \rightarrow hh \rightarrow b\bar{b}\tau^{+}\tau^{-})$ (upper panel) and $\sigma(pp \rightarrow hh \rightarrow b\bar{b}\gamma\gamma)$ (lower panel) normalized to SM value as a function of the SM-like Higgs boson production cross section via gluon-gluon fusion, with the SM-like Higgs boson decaying into $\gamma\gamma$ pair. All points satisfy current experimental constraints. The horizontal line is the 95\% CL upper limit of the projected di-Higgs production cross section at HL-LHC. The vertical lines are the 95\% CL limits of the projected $\sigma(ggF,h \rightarrow \gamma\gamma)$ at HL-LHC. The magenta dots are outside the 95\% CL of the projected SM-like Higgs boson production cross sections via other production modes and decay modes at HL-LHC. The blue dots are outside the 95\% CL of the projected search for the additional neutral Higgs Bosons in the di-tau final state at HL-LHC.}
\end{figure}

In the NMSSM study, 
	we perform a random scan over the NMSSM parameter space using \textbf{NMSSMTools-5.4.1} as there are more relevant parameters. A similar study can be found in ~\cite{Basler:2018dac}. The range of the parameters is as shown in Tab.\ref{tab:1}. We also assume 
\begin{equation}
m_{\tilde{t}_{R}} = m_{\tilde{Q}_{3}} \,,\quad m_{\tilde{\tau}_{R}} = m_{\tilde{L}_{3}} \,,\quad m_{\tilde{b}_{R}} = 3 \,\mathrm{TeV} \,,
\end{equation}
although we expect the sfermion contributions are negligible, as in the MSSM case. We require the mass of one of the NMSSM neutral CP-even Higgs bosons to be 125.26$\pm$3~GeV to accommodate experimental and theoretical uncertainties up to 3 GeV. 
Besides that, $\lambda$ and $\kappa$ have to be sufficiently small so that perturbation theory remains valid. Here, we use~\cite{Carena:2015moc}
\begin{equation}\label{eq:10}
\lambda^{2} + \kappa^{2} < 0.7^{2} \,.
\end{equation}  
Constraints from collider experiments (LEP, Tevatron, LHC) and dark matter direct detection experiments are also checked by using \textbf{NMSSMTools-5.4.1}~\cite{Belanger:2004yn,Belanger:2006is,Belanger:2013oya,NT04,NT05,NT09,NT97,NT07,NT05DM}. All points in Fig.\ref{fig:6} also satisfy the constraint on dark matter relic density from \textit{Planck} measurement~\cite{planck18} (including +10\% uncertainty in theoretical calculation) 
\begin{equation}\label{eq:11}
\Omega_{\tilde{\chi}_{1}^{0}}h^{2} \leq 0.131 \,.
\end{equation}

\begin{table}[tb]
  	\raggedright
  	\resizebox{\textwidth}{0.055\textheight}{\begin{minipage}{\textwidth}
  	\centering
  	\begin{tabular}{|c|ccc|ccccccccccc|}
  	\hline
  		& tan$\beta$ & $\lambda$ & $\kappa$	& $A_{\lambda}$	& $A_{\kappa}$	& $\mu_{\mathrm{eff}}$ & $M_{1}$	& $M_{2}$ &	$M_{3}$ & $A_{t}$ &	$A_{b}$ & $A_{\tau}$& $m_{\tilde{Q}_3}$ & $m_{\tilde{L}_3}$	\\
  		&			 &		   	 &			& \multicolumn{11}{|c|}{(in TeV)}		\\  
	\hline
  	min	& 1			 &	0		 & -0.7		& -1		   	& -1			& -0.5		  &	0.1		& 0.2	  &	1.3		& -6	  &	-6		& -3		& 0.6	   	  &	0.6			 \\
  		& 			 &			 &			&		   		&				&			  &			&		  &			&		  &			&			&		   	  &				 \\[-1em]  
	\hline
  	max	& 10		 &	0.7		 & 0.7		& 1		   		& 1				& 0.5		  &	1		& 2		  &	7		& 6		  &	6		& 3			& 4		   	  &	4			 \\
	\hline
    \end{tabular}
	\end{minipage} }
	\caption{\label{tab:1} NMSSM parameter space.}
 	\end{table}
 
\par
	Our results for the NMSSM are shown in Fig.\ref{fig:6}, for each points surviving all constraints, we plot the di-Higgs production cross section in the $bb\tau^{+}\tau^{-}$ channel (upper panel) and the $bb\gamma\gamma$ channel (lower panel).
	In the region of interest, $B(h \rightarrow b\bar{b})B(h \rightarrow \tau^{+}\tau^{-})$ can be enhanced up to 16\% compared to the SM value, and $B(h \rightarrow b\bar{b})B(h \rightarrow \gamma\gamma)$ can be suppressed up to 10\% compared to the SM value.
	The NMSSM di-Higgs production cross sections can be enhanced significantly compared to the SM. Similar to the MSSM case, the interference between the resonant and non-resonant amplitude and the contributions from squark loops are small over the whole parameter space. When the new CP-even Higgs states are light, the dominant contribution is from resonant di-Higgs production. The resonant di-Higgs rate can be a few times the SM rate. The non-resonant contribution of di-Higgs production is generally enhanced by 20\% to 60\%, which comes from the decrease in the SM-like Higgs self-coupling. In the region of interest, $\delta_{3}$ generally varies between -0.3 and -0.58, resulting in the 20$\%$ to 60 $\%$ enhancement in the non-resonant production. Here we use the projected sensitivity of the non-resonant production, and we expect it to be better for resonant productions.
	Unlike the MSSM case, $\kappa_{t}$ is very close to unity in the NMSSM case, and therefore, the enhancement due to the suppression in $\lambda_3$ is not offset by the suppression in $\kappa_t$. In most of the parameter space, the di-Higgs rate is enhanced through the suppression in $\lambda_3$.   
	
	\par The horizontal lines in Fig.\ref{fig:6} show the projected sensitivity at the HL-LHC for the $bb\tau\tau$ channel (upper panel), and the $bb\gamma\gamma$ channel (lower panel). Most points lie below the horizontal lines, meaning the HL-LHC di-Higgs production measurements have very limited sensitivity in most regions of interest.
	
	\par We also study the complementarity between di-Higgs and other Higgs measurements. In terms of precision Higgs measurements, we found that $h \rightarrow \gamma\gamma$ from gluon fusion is the most sensitive channel among all Higgs boson production modes and decay channels at the HL-LHC. To compare with precision Higgs measurements, we plot the $h\rightarrow \gamma\gamma$ cross section from gluon fusion and overlay the projected sensitivity in this channel as the vertical lines at the HL-LHC. The $h\rightarrow\gamma\gamma$ measurement can probe a large fraction of the parameter space, and almost all points that can be tested by the di-Higgs measurement, can be tested by the $h\rightarrow\gamma\gamma$ measurement. To consider the sensitivity in precision Higgs measurements, we use magenta dots to represent points that can be tested by other Higgs coupling measurements. Most points that can be tested by the $h\rightarrow\gamma\gamma$ measurement can be probed by other Higgs measurements as well, such as $h\rightarrow Z Z$ and $h\rightarrow W W$ measurements. 
\par  For direct searches of the heavy Higgs, we found that the search for heavy neutral Higgs bosons in the di-tau final state at HL-LHC~\cite{HHditauAtlasHL} is a good complementary probe. We use blue dots to represent the points that can be tested in the heavy Higgs to di-tau channel. From Fig.\ref{fig:6}, we can see that many blue dots lie between the two vertical lines, and below the horizontal line, showing direct searches of a heavy Higgs provides a complementary probe to precision Higgs and di-Higgs measurements. Combining di-Higgs, precision Higgs, and direct searches of the heavy Higgs, a large set of parameter points can be tested at the HL-LHC. 

\section{Conclusions}
In this work, we have calculated the di-Higgs production cross section in the MSSM and the NMSSM at the one loop level. We include the possible resonant contribution from new CP-even neutral Higgs states, new colored particles, and possible modifications in the Higgs couplings. We found that in both cases, the di-Higgs production can be enhanced significantly through a new resonance. As the new CP-even neutral Higgs bosons become heavy, di-Higgs production only enhances moderately due to the suppression in the Higgs self-couplings. The di-Higgs rate in the $bb\tau^{+}\tau^{-}$ final state can be further enhanced through the enhanced branching ratios. Given the current stop limit, we found the contributions from stops and other sfermions are small. Also, the interference effect is found to be small.

\par We further study the complementarity of di-Higgs measurement to other Higgs studies, including the precision measurement of the Higgs couplings and the search for new Higgs bosons. We found that there is a strong correlation in di-Higgs production rate and single Higgs production rate, especially outside the region where resonant di-Higgs dominates. Due to the complexity of the signature, and the small production cross section, di-Higgs measurements are not as sensitive as single Higgs measurements. The direct searches to new Higgs bosons, on the other hand, provide a complementary probe.

\acknowledgments
We thank I. Lewis, K. Hagowara, and S. Heinemeyer for helpful discussions. This work is supported by University of Nebraska-Lincoln, National Science Foundation under grant number PHY-1820891, and the NSF Nebraska EPSCoR under grant number OIA-1557417.
\newpage
\appendix
\section{Form Factors}\label{sec:form}
For the partonic process $g(p_{1})g(p_{2}) \rightarrow h(k_{1})h(k_{2})$,
\\
\begin{align}
F_{\triangleright}^{(1/2)}(m_{f}) &= \dfrac{2m_f^2}{\hat{s}}[2 + (4 m_f^2 - \hat{s}) C_0(0,0,\hat{s},m_f^2,m_f^2,m_f^2)] \,,
\\
F_{\square}^{(1/2)}(m_{f}) &= \dfrac{2m_f^2}{\hat{s}}\big\{m_f^2(8 m_f^2 -\hat{s} - 2 m_h^2)  \nonumber
\\
&[D_0(0,0,m_h^2,m_h^2,\hat{s},\hat{u},m_f^2,m_f^2,m_f^2,m_f^2) + D_0(0,0,m_h^2,m_h^2,\hat{s},\hat{t},m_f^2,m_f^2,m_f^2,m_f^2) \nonumber
\\
&+ D_0(0,m_h^2,0,m_h^2,\hat{t},\hat{u},m_f^2,m_f^2,m_f^2,m_f^2)] \nonumber
\\
&+ \dfrac{\hat{u}\hat{t}-m_h^4}{\hat{s}}(4 m_f^2-m_h^2)D_0(0,m_h^2,0,m_h^2,\hat{t},\hat{u},m_f^2,m_f^2,m_f^2,m_f^2) \nonumber
\\
&+ 2 + 4 m_f^2 C_0(0,0,\hat{s},m_f^2,m_f^2,m_f^2) \nonumber
\\
&+ \dfrac{2}{\hat{s}}(m_h^2-4 m_f^2)[\,(\hat{t}-m_h^2)C_0(0,m_h^2,\hat{t},m_f^2,m_f^2,m_f^2) \nonumber
\\
&\qquad\qquad\qquad\quad        + (\hat{u}-m_h^2)C_0(0,m_h^2,\hat{u},m_f^2,m_f^2,m_f^2)\,]\, \big\} \,,
\\
G_{\square}^{(1/2)}(m_{f}) &= \dfrac{m_f^2}{\hat{s}}\bigg\{ 2(8 m_f^2+\hat{s}-2m_h^2) \nonumber
\\
& \big\{ m_f^2[D_0(0,0,m_h^2,m_h^2,\hat{s},\hat{u},m_f^2,m_f^2,m_f^2,m_f^2) + D_0(0,0,m_h^2,m_h^2,\hat{s},\hat{t},m_f^2,m_f^2,m_f^2,m_f^2) \nonumber
\\
&+ D_0(0,m_h^2,0,m_h^2,\hat{t},\hat{u},m_f^2,m_f^2,m_f^2,m_f^2)]-C_0(m_h^2, m_h^2,\hat{s}, m_f^2,m_f^2,m_f^2) \big\} \nonumber
\\
& -2\big\{ \hat{s}C_0(0,0,\hat{s},m_f^2,m_f^2,m_f^2) + (\hat{t} - m_h^2)C_0(0,m_h^2,\hat{t},m_f^2,m_f^2,m_f^2) \nonumber
\\
&\qquad + (\hat{u} - m_h^2)C_0(0,m_h^2,\hat{u},m_f^2,m_f^2,m_f^2) \big\} \nonumber
\\
& +\dfrac{1}{ut-mh^4} \Big\{ \hat{s}\hat{u}(8 \hat{u}m_f^2 - u^2 - m_h^4)D_0(0,0,m_h^2,m_h^2,\hat{s},\hat{u},m_f^2,m_f^2,m_f^2,m_f^2) \nonumber
\\
&+ \hat{s}\hat{t}(8 \hat{t}m_f^2-\hat{t}^2-m_h^4)D_0(0,0,m_h^2,m_h^2,\hat{s},\hat{t},m_f^2,m_f^2,m_f^2,m_f^2) \nonumber
\\
&+ (8 m_f^2+\hat{s}-2m_h^2) \big\{ \hat{s}(\hat{s} - 2m_h^2) C_0(0,0,\hat{s},m_f^2,m_f^2,m_f^2) \nonumber
\\
&\qquad\qquad\qquad\quad + \hat{s}(\hat{s}-4m_h^2)C_0(m_h^2,m_h^2,\hat{s},m_f^2,m_f^2,m_f^2) \nonumber
\\
&\qquad\qquad\qquad\quad + 2\hat{t}(m_h^2 - \hat{t})C_0(0,m_h^2,\hat{t},m_f^2,m_f^2,m_f^2) \nonumber
\\
&\qquad\qquad\qquad\quad + 2\hat{u}(m_h^2 - \hat{u})C_0(0,m_h^2,\hat{u},m_f^2,m_f^2,m_f^2) \big\} \,\Big\} \bigg\}
\\
F_{\triangleright}^{(0)}(m_{s}) &= \dfrac{-2m_s^2}{\hat{s}}(1 + 2 m_s^2 C_0(0,0,\hat{s},m_f^2,m_f^2,m_f^2))
\\
F_{\square}^{(0)}(m_{s},m_{s'}) &= \dfrac{-4m_s^2m_{s'}^2}{\hat{s}}\Big\{\dfrac{m_h^2-\hat{t}}{2\hat{s}}[C_0(m_h^2, 0, \hat{t}, m_{s'}^2, m_s^2, m_s^2) + C_0(m_h^2, 0, \hat{t}, m_s^2, m_{s'}^2, m_{s'}^2)] \nonumber
\\
&+ \dfrac{m_h^2-\hat{u}}{2\hat{s}}[C_0(m_h^2, 0, \hat{u}, m_{s'}^2, m_s^2, m_s^2) + C_0(m_h^2, 0, \hat{u}, m_s^2, m_{s'}^2, m_{s'}^2)]  \nonumber
\\
&+ \dfrac{1}{2}(m_{s'}^2-m_s^2 + \dfrac{\hat{u}\hat{t} - m_h^4}{\hat{s}})D_0(m_h^2, 0, m_h^2, 0, \hat{t}, \hat{u},m_{s'}^2,m_s^2, m_s^2, m_{s'}^2)  \nonumber
\\
&+ m_s^2 [D_0(m_h^2, 0, m_h^2, 0, \hat{t},\hat{u}, m_{s'}^2, m_s^2, m_s^2, m_{s'}^2) + D_0(m_h^2, m_h^2, 0, 0, \hat{s}, \hat{t}, m_s^2, m_{s'}^2, m_s^2, m_s^2) \nonumber
\\
&\qquad\quad + D_0(m_h^2, m_h^2, 0, 0, \hat{s}, \hat{u}, m_s^2, m_{s'}^2, m_s^2, m_s^2)] \,\Big\}
\end{align}
\begin{align}
G_{\square}^{(0)}(m_{s},m_{s'}) &= \dfrac{-2 m_s^2 m_{s'}^2}{\hat{s}(m_h^4-\hat{u}\hat{t})} \Big\{ \hat{s} (2m_s^2-2m_{s'}^2+\hat{t}+\hat{u})C_0(0, 0, s, m_s^2, m_s^2, m_s^2) \nonumber
\\
&-2\hat{t}(m_h^2 - \hat{t})C_0(m_h^2, 0, \hat{t}, m_{s'}^2, m_s^2, m_s^2) - 2\hat{u}(m_h^2 - \hat{u})C_0(m_h^2, 0, \hat{u}, m_{s'}^2, m_s^2, m_s^2) \nonumber
\\ 
&- (m_h^2-\hat{t})(m_s^2-m_{s'}^2)[C_0(m_h^2, 0, \hat{t}, m_s^2, m_{s'}^2, m_{s'}^2) + C_0(m_h^2, 0, \hat{t}, m_{s'}^2, m_s^2, m_s^2)] \nonumber
\\
&- (m_h^2 -\hat{u})(m_s^2-m_{s'}^2)[C_0(m_h^2, 0, \hat{u}, m_s^2, m_{s'}^2, m_{s'}^2) + C_0(m_h^2, 0, \hat{u}, m_{s'}^2, m_s^2, m_s^2)] \nonumber
\\
&+ (2 m_h^4 -\hat{t}^2 - \hat{u}^2) C_0(m_h^2, m_h^2, \hat{s},m_s^2, m_{s'}^2, m_s^2) \nonumber
\\
&+ [-\hat{s} (m_s^2 - m_{s'}^2)^2 + (m_s^2 + m_{s'}^2) (m_h^4 -\hat{u}\hat{t})][D_0(m_h^2, 0, m_h^2, 0, \hat{t},\hat{u}, m_{s'}^2, m_s^2, m_s^2, m_{s'}^2) \nonumber
\\
&+ D_0(m_h^2, m_h^2, 0, 0, \hat{s}, \hat{t}, m_s^2, m_{s'}^2, m_s^2, m_s^2) + D_0(m_h^2, m_h^2, 0, 0, \hat{s}, \hat{u}, m_s^2, m_{s'}^2, m_s^2, m_s^2)] \nonumber
\\
&- [\hat{s}\hat{t}^2+(m_s^2-m_{s'}^2)(2\hat{s}\hat{t}+\hat{u}\hat{t}-m_h^4)] D_0(m_h^2, m_h^2, 0, 0, \hat{s}, \hat{t}, m_s^2, m_{s'}^2, m_s^2, m_s^2) \nonumber
\\
&- [\hat{s}\hat{u}^2+(m_s^2-m_{s'}^2)(2\hat{s}\hat{u}+\hat{u}\hat{t}-m_h^4)] D_0(m_h^2,m_h^2,0,0,\hat{s},\hat{u}, m_s^2, m_{s'}^2, m_s^2, m_s^2) \Big\}
\end{align}

where
\begin{align}
\hat{s} &= (p_1 + p_2)^2 \,,
\\
\hat{t} &= (p_1 - k_1)^2 \,,
\\
\hat{u} &= (p_1 - k_2)^2 \,,
\end{align}
and $C_0$ and $D_0$ are scalar three-point and four-point functions of one-loop integrals, defined as
\begin{align}
&C_0 (p_1^2,p_2^2,(p_1+p_2)^2,m_1^2,m_2^2,m_3^2) \nonumber
\\
&= \dfrac{\mu^{4-D}}{i\pi^{D/2}r_{\Gamma}} \int \dfrac{d^D q}{[q^2-m_1^2][(q+p_1)^2-m_2^2][(q+p_1+p_2)^2-m_3^2]} \,,
\\
&D_0 (p_1^2,p_2^2,p_3^2,p_4^2,(p_1+p_2)^2,(p_2+p_3)^2,m_1^2,m_2^2,m_3^2,m_4^2) \nonumber
\\
&= \dfrac{\mu^{4-D}}{i\pi^{D/2}r_{\Gamma}} \int \dfrac{d^D q}{[q^2-m_1^2][(q+p_1)^2-m_2^2][(q+p_1+p_2)^2-m_3^2][(q+p_1+p_2+p_3)^2-m_4^2]} \,,
\end{align}
where
\begin{equation}
r_{\Gamma} = \dfrac{\Gamma^2 (1-\epsilon)\Gamma (1+\epsilon)}{\Gamma (1-2\epsilon)} \,,\quad D = 4-2\epsilon \,.
\end{equation}
The integrals are independent of $\mu$ in the limit $\epsilon \rightarrow 0$. We follow the conventions for momenta as stated in the manual of \textbf{LoopTools-2.15}~\cite{LTHahn1998}.

\bibliographystyle{JHEP}
\bibliography{refList}

\providecommand{\href}[2]{#2}\begingroup\raggedright\begin{thebibliography}{10}

\bibitem{higgs2012Atlas}
{\scshape ATLAS} collaboration, G.~Aad et~al., \emph{{Observation of a new
  particle in the search for the Standard Model Higgs boson with the ATLAS
  detector at the LHC}},
  \href{https://doi.org/10.1016/j.physletb.2012.08.020}{\emph{Phys. Lett.}
  {\bfseries B716} (2012) 1} [\href{https://arxiv.org/abs/1207.7214}{{\ttfamily
  1207.7214}}].

\bibitem{higgs2012CMS}
{\scshape CMS} collaboration, S.~Chatrchyan et~al., \emph{{Observation of a New
  Boson at a Mass of 125 GeV with the CMS Experiment at the LHC}},
  \href{https://doi.org/10.1016/j.physletb.2012.08.021}{\emph{Phys. Lett.}
  {\bfseries B716} (2012) 30}
  [\href{https://arxiv.org/abs/1207.7235}{{\ttfamily 1207.7235}}].

\bibitem{higgsSPCMS}
{\scshape CMS} collaboration, V.~Khachatryan et~al., \emph{{Constraints on the
  spin-parity and anomalous HVV couplings of the Higgs boson in proton
  collisions at 7 and 8 TeV}},
  \href{https://doi.org/10.1103/PhysRevD.92.012004}{\emph{Phys. Rev.}
  {\bfseries D92} (2015) 012004}
  [\href{https://arxiv.org/abs/1411.3441}{{\ttfamily 1411.3441}}].

\bibitem{higgsSPAtlas}
{\scshape ATLAS} collaboration, G.~Aad et~al., \emph{{Study of the spin and
  parity of the Higgs boson in diboson decays with the ATLAS detector}},
  \href{https://doi.org/10.1140/epjc/s10052-015-3685-1,
  10.1140/epjc/s10052-016-3934-y}{\emph{Eur. Phys. J.} {\bfseries C75} (2015)
  476} [\href{https://arxiv.org/abs/1506.05669}{{\ttfamily 1506.05669}}].

\bibitem{HcouplingsCMS}
{\scshape CMS} collaboration, A.~M. Sirunyan et~al., \emph{{Combined
  measurements of Higgs boson couplings in proton–proton collisions at
  $\sqrt{s}=13\,\text {Te}\text {V} $}},
  \href{https://doi.org/10.1140/epjc/s10052-019-6909-y}{\emph{Eur. Phys. J.}
  {\bfseries C79} (2019) 421}
  [\href{https://arxiv.org/abs/1809.10733}{{\ttfamily 1809.10733}}].

\bibitem{HcouplingsAtlas}
{\scshape ATLAS} collaboration, \emph{{Combined measurements of Higgs boson
  production and decay using up to $80$ fb$^{-1}$ of proton--proton collision
  data at $\sqrt{s}=$ 13 TeV collected with the ATLAS experiment}},  Tech. Rep.
  ATLAS-CONF-2019-005, CERN, Geneva, Mar, 2019.

\bibitem{Chen:2014ask}
C.-Y. Chen, S.~Dawson and I.~M. Lewis, \emph{{Exploring resonant di-Higgs boson
  production in the Higgs singlet model}},
  \href{https://doi.org/10.1103/PhysRevD.91.035015}{\emph{Phys. Rev.}
  {\bfseries D91} (2015) 035015}
  [\href{https://arxiv.org/abs/1410.5488}{{\ttfamily 1410.5488}}].

\bibitem{Huang:2015tdv}
P.~Huang, A.~Joglekar, B.~Li and C.~E.~M. Wagner, \emph{{Probing the
  Electroweak Phase Transition at the LHC}},
  \href{https://doi.org/10.1103/PhysRevD.93.055049}{\emph{Phys. Rev.}
  {\bfseries D93} (2016) 055049}
  [\href{https://arxiv.org/abs/1512.00068}{{\ttfamily 1512.00068}}].

\bibitem{Lewis:2017dme}
I.~M. Lewis and M.~Sullivan, \emph{{Benchmarks for Double Higgs Production in
  the Singlet Extended Standard Model at the LHC}},
  \href{https://doi.org/10.1103/PhysRevD.96.035037}{\emph{Phys. Rev.}
  {\bfseries D96} (2017) 035037}
  [\href{https://arxiv.org/abs/1701.08774}{{\ttfamily 1701.08774}}].

\bibitem{DiMicco:2019ngk}
J.~Alison et~al., \emph{{Higgs Boson Pair Production at Colliders: Status and
  Perspectives}},  in \emph{{Double Higgs Production at Colliders Batavia, IL,
  USA, September 4, 2018-9, 2019}} (B.~Di~Micco, M.~Gouzevitch, J.~Mazzitelli
  and C.~Vernieri, eds.), 2019,
  \href{https://arxiv.org/abs/1910.00012}{{\ttfamily 1910.00012}},
  \href{https://lss.fnal.gov/archive/2019/conf/fermilab-conf-19-468-e-t.pdf}{https://lss.fnal.gov/archive/2019/conf/fermilab-conf-19-468-e-t.pdf}.

\bibitem{Batell:2015koa}
B.~Batell, M.~McCullough, D.~Stolarski and C.~B. Verhaaren, \emph{{Putting a
  Stop to di-Higgs Modifications}},
  \href{https://doi.org/10.1007/JHEP09(2015)216}{\emph{JHEP} {\bfseries 09}
  (2015) 216} [\href{https://arxiv.org/abs/1508.01208}{{\ttfamily
  1508.01208}}].

\bibitem{Dawson:2015oha}
S.~Dawson, A.~Ismail and I.~Low, \emph{{What’s in the loop? The anatomy of
  double Higgs production}},
  \href{https://doi.org/10.1103/PhysRevD.91.115008}{\emph{Phys. Rev.}
  {\bfseries D91} (2015) 115008}
  [\href{https://arxiv.org/abs/1504.05596}{{\ttfamily 1504.05596}}].

\bibitem{Huang:2017nnw}
P.~Huang, A.~Joglekar, M.~Li and C.~E.~M. Wagner, \emph{{Corrections to
  di-Higgs boson production with light stops and modified Higgs couplings}},
  \href{https://doi.org/10.1103/PhysRevD.97.075001}{\emph{Phys. Rev.}
  {\bfseries D97} (2018) 075001}
  [\href{https://arxiv.org/abs/1711.05743}{{\ttfamily 1711.05743}}].

\bibitem{Chen:2017qcz}
C.-Y. Chen, J.~Kozaczuk and I.~M. Lewis, \emph{{Non-resonant Collider
  Signatures of a Singlet-Driven Electroweak Phase Transition}},
  \href{https://doi.org/10.1007/JHEP08(2017)096}{\emph{JHEP} {\bfseries 08}
  (2017) 096} [\href{https://arxiv.org/abs/1704.05844}{{\ttfamily
  1704.05844}}].

\bibitem{Basler:2019nas}
P.~Basler, S.~Dawson, C.~Englert and M.~Mühlleitner, \emph{{Di-Higgs Peaks and
  Top Valleys: Interference Effects in Higgs Sector Extensions}},
  \href{https://arxiv.org/abs/1909.09987}{{\ttfamily 1909.09987}}.

\bibitem{Carena:2018vpt}
M.~Carena, Z.~Liu and M.~Riembau, \emph{{Probing the electroweak phase
  transition via enhanced di-Higgs boson production}},
  \href{https://doi.org/10.1103/PhysRevD.97.095032}{\emph{Phys. Rev.}
  {\bfseries D97} (2018) 095032}
  [\href{https://arxiv.org/abs/1801.00794}{{\ttfamily 1801.00794}}].

\bibitem{FH98_1}
S.~Heinemeyer, W.~Hollik and G.~Weiglein, \emph{{FeynHiggs: A Program for the
  calculation of the masses of the neutral CP even Higgs bosons in the MSSM}},
  \href{https://doi.org/10.1016/S0010-4655(99)00364-1}{\emph{Comput. Phys.
  Commun.} {\bfseries 124} (2000) 76}
  [\href{https://arxiv.org/abs/hep-ph/9812320}{{\ttfamily hep-ph/9812320}}].

\bibitem{FH98_2}
S.~Heinemeyer, W.~Hollik and G.~Weiglein, \emph{{The Masses of the neutral CP -
  even Higgs bosons in the MSSM: Accurate analysis at the two loop level}},
  \href{https://doi.org/10.1007/s100529900006}{\emph{Eur. Phys. J.} {\bfseries
  C9} (1999) 343} [\href{https://arxiv.org/abs/hep-ph/9812472}{{\ttfamily
  hep-ph/9812472}}].

\bibitem{FH02}
G.~Degrassi, S.~Heinemeyer, W.~Hollik, P.~Slavich and G.~Weiglein,
  \emph{{Towards high precision predictions for the MSSM Higgs sector}},
  \href{https://doi.org/10.1140/epjc/s2003-01152-2}{\emph{Eur. Phys. J.}
  {\bfseries C28} (2003) 133}
  [\href{https://arxiv.org/abs/hep-ph/0212020}{{\ttfamily hep-ph/0212020}}].

\bibitem{FH06}
M.~Frank, T.~Hahn, S.~Heinemeyer, W.~Hollik, H.~Rzehak and G.~Weiglein,
  \emph{{The Higgs Boson Masses and Mixings of the Complex MSSM in the
  Feynman-Diagrammatic Approach}},
  \href{https://doi.org/10.1088/1126-6708/2007/02/047}{\emph{JHEP} {\bfseries
  02} (2007) 047} [\href{https://arxiv.org/abs/hep-ph/0611326}{{\ttfamily
  hep-ph/0611326}}].

\bibitem{FH13}
T.~Hahn, S.~Heinemeyer, W.~Hollik, H.~Rzehak and G.~Weiglein,
  \emph{{High-Precision Predictions for the Light CP -Even Higgs Boson Mass of
  the Minimal Supersymmetric Standard Model}},
  \href{https://doi.org/10.1103/PhysRevLett.112.141801}{\emph{Phys. Rev. Lett.}
  {\bfseries 112} (2014) 141801}
  [\href{https://arxiv.org/abs/1312.4937}{{\ttfamily 1312.4937}}].

\bibitem{FH16}
H.~Bahl and W.~Hollik, \emph{{Precise prediction for the light MSSM Higgs boson
  mass combining effective field theory and fixed-order calculations}},
  \href{https://doi.org/10.1140/epjc/s10052-016-4354-8}{\emph{Eur. Phys. J.}
  {\bfseries C76} (2016) 499}
  [\href{https://arxiv.org/abs/1608.01880}{{\ttfamily 1608.01880}}].

\bibitem{FH17}
H.~Bahl, S.~Heinemeyer, W.~Hollik and G.~Weiglein, \emph{{Reconciling EFT and
  hybrid calculations of the light MSSM Higgs-boson mass}},
  \href{https://doi.org/10.1140/epjc/s10052-018-5544-3}{\emph{Eur. Phys. J.}
  {\bfseries C78} (2018) 57}
  [\href{https://arxiv.org/abs/1706.00346}{{\ttfamily 1706.00346}}].

\bibitem{Campbell2006wx}
J.~M. Campbell, J.~W. Huston and W.~J. Stirling, \emph{{Hard Interactions of
  Quarks and Gluons: A Primer for LHC Physics}},
  \href{https://doi.org/10.1088/0034-4885/70/1/R02}{\emph{Rept. Prog. Phys.}
  {\bfseries 70} (2007) 89}
  [\href{https://arxiv.org/abs/hep-ph/0611148}{{\ttfamily hep-ph/0611148}}].

\bibitem{Ellwanger2009dp}
U.~Ellwanger, C.~Hugonie and A.~M. Teixeira, \emph{{The Next-to-Minimal
  Supersymmetric Standard Model}},
  \href{https://doi.org/10.1016/j.physrep.2010.07.001}{\emph{Phys. Rept.}
  {\bfseries 496} (2010) 1} [\href{https://arxiv.org/abs/0910.1785}{{\ttfamily
  0910.1785}}].

\bibitem{NT04}
U.~Ellwanger, J.~F. Gunion and C.~Hugonie, \emph{{NMHDECAY: A Fortran code for
  the Higgs masses, couplings and decay widths in the NMSSM}},
  \href{https://doi.org/10.1088/1126-6708/2005/02/066}{\emph{JHEP} {\bfseries
  02} (2005) 066} [\href{https://arxiv.org/abs/hep-ph/0406215}{{\ttfamily
  hep-ph/0406215}}].

\bibitem{NT05}
U.~Ellwanger and C.~Hugonie, \emph{{NMHDECAY 2.0: An Updated program for
  sparticle masses, Higgs masses, couplings and decay widths in the NMSSM}},
  \href{https://doi.org/10.1016/j.cpc.2006.04.004}{\emph{Comput. Phys. Commun.}
  {\bfseries 175} (2006) 290}
  [\href{https://arxiv.org/abs/hep-ph/0508022}{{\ttfamily hep-ph/0508022}}].

\bibitem{NT09}
G.~Degrassi and P.~Slavich, \emph{{On the radiative corrections to the neutral
  Higgs boson masses in the NMSSM}},
  \href{https://doi.org/10.1016/j.nuclphysb.2009.09.018}{\emph{Nucl. Phys.}
  {\bfseries B825} (2010) 119}
  [\href{https://arxiv.org/abs/0907.4682}{{\ttfamily 0907.4682}}].

\bibitem{NT97}
A.~Djouadi, J.~Kalinowski and M.~Spira, \emph{{HDECAY: A Program for Higgs
  boson decays in the standard model and its supersymmetric extension}},
  \href{https://doi.org/10.1016/S0010-4655(97)00123-9}{\emph{Comput. Phys.
  Commun.} {\bfseries 108} (1998) 56}
  [\href{https://arxiv.org/abs/hep-ph/9704448}{{\ttfamily hep-ph/9704448}}].

\bibitem{NT07}
F.~Domingo and U.~Ellwanger, \emph{{Updated Constraints from $B$ Physics on the
  MSSM and the NMSSM}},
  \href{https://doi.org/10.1088/1126-6708/2007/12/090}{\emph{JHEP} {\bfseries
  12} (2007) 090} [\href{https://arxiv.org/abs/0710.3714}{{\ttfamily
  0710.3714}}].

\bibitem{NT05DM}
G.~Belanger, F.~Boudjema, C.~Hugonie, A.~Pukhov and A.~Semenov, \emph{{Relic
  density of dark matter in the NMSSM}},
  \href{https://doi.org/10.1088/1475-7516/2005/09/001}{\emph{JCAP} {\bfseries
  0509} (2005) 001} [\href{https://arxiv.org/abs/hep-ph/0505142}{{\ttfamily
  hep-ph/0505142}}].

\bibitem{mstAtlas}
{\scshape ATLAS} collaboration, \emph{{Search for direct top squark pair
  production in the 3-body decay mode with a final state containing one lepton,
  jets, and missing transverse momentum in $\sqrt{s}=13$TeV $pp$ collision data
  with the ATLAS detector}},  Tech. Rep. ATLAS-CONF-2019-017, CERN, Geneva,
  May, 2019.

\bibitem{mstCMS}
{\scshape CMS} collaboration, \emph{{Search for direct top squark pair
  production in events with one lepton, jets and missing transverse energy at
  13 TeV}},  Tech. Rep. CMS-PAS-SUS-19-009, CERN, Geneva, 2019.

\bibitem{vacstabbound}
N.~Blinov and D.~E. Morrissey, \emph{{Vacuum Stability and the MSSM Higgs
  Mass}}, \href{https://doi.org/10.1007/JHEP03(2014)106}{\emph{JHEP} {\bfseries
  03} (2014) 106} [\href{https://arxiv.org/abs/1310.4174}{{\ttfamily
  1310.4174}}].

\bibitem{HHditauCMS}
{\scshape CMS} collaboration, A.~M. Sirunyan et~al., \emph{{Search for
  additional neutral MSSM Higgs bosons in the $\tau\tau$ final state in
  proton-proton collisions at $\sqrt{s}=$ 13 TeV}},
  \href{https://doi.org/10.1007/JHEP09(2018)007}{\emph{JHEP} {\bfseries 09}
  (2018) 007} [\href{https://arxiv.org/abs/1803.06553}{{\ttfamily
  1803.06553}}].

\bibitem{HHditauAtlas}
{\scshape ATLAS} collaboration, M.~Aaboud et~al., \emph{{Search for additional
  heavy neutral Higgs and gauge bosons in the ditau final state produced in 36
  fb$^{−1}$ of pp collisions at $\sqrt{s}=$ 13 TeV with the ATLAS detector}},
  \href{https://doi.org/10.1007/JHEP01(2018)055}{\emph{JHEP} {\bfseries 01}
  (2018) 055} [\href{https://arxiv.org/abs/1709.07242}{{\ttfamily
  1709.07242}}].

\bibitem{HHditauAtlasHL}
{\scshape ATLAS} collaboration, \emph{{Prospects for the search for additional
  Higgs bosons in the ditau final state with the ATLAS detector at HL-LHC}},
  Tech. Rep. ATL-PHYS-PUB-2018-050, CERN, Geneva, Dec, 2018.

\bibitem{HcouplingsAtlasHL}
{\scshape ATLAS} collaboration, \emph{{Projections for measurements of Higgs
  boson cross sections, branching ratios, coupling parameters and mass with the
  ATLAS detector at the HL-LHC}},  Tech. Rep. ATL-PHYS-PUB-2018-054, CERN,
  Geneva, Dec, 2018.

\bibitem{Basler:2018dac}
P.~Basler, S.~Dawson, C.~Englert and M.~Mühlleitner, \emph{{Showcasing HH
  production: Benchmarks for the LHC and HL-LHC}},
  \href{https://doi.org/10.1103/PhysRevD.99.055048}{\emph{Phys. Rev.}
  {\bfseries D99} (2019) 055048}
  [\href{https://arxiv.org/abs/1812.03542}{{\ttfamily 1812.03542}}].

\bibitem{Carena:2015moc}
M.~Carena, H.~E. Haber, I.~Low, N.~R. Shah and C.~E.~M. Wagner,
  \emph{{Alignment limit of the NMSSM Higgs sector}},
  \href{https://doi.org/10.1103/PhysRevD.93.035013}{\emph{Phys. Rev.}
  {\bfseries D93} (2016) 035013}
  [\href{https://arxiv.org/abs/1510.09137}{{\ttfamily 1510.09137}}].

\bibitem{Belanger:2004yn}
G.~Belanger, F.~Boudjema, A.~Pukhov and A.~Semenov, \emph{{micrOMEGAs: Version
  1.3}}, \href{https://doi.org/10.1016/j.cpc.2005.12.005}{\emph{Comput. Phys.
  Commun.} {\bfseries 174} (2006) 577}
  [\href{https://arxiv.org/abs/hep-ph/0405253}{{\ttfamily hep-ph/0405253}}].

\bibitem{Belanger:2006is}
G.~Belanger, F.~Boudjema, A.~Pukhov and A.~Semenov, \emph{{MicrOMEGAs 2.0: A
  Program to calculate the relic density of dark matter in a generic model}},
  \href{https://doi.org/10.1016/j.cpc.2006.11.008}{\emph{Comput. Phys. Commun.}
  {\bfseries 176} (2007) 367}
  [\href{https://arxiv.org/abs/hep-ph/0607059}{{\ttfamily hep-ph/0607059}}].

\bibitem{Belanger:2013oya}
G.~Belanger, F.~Boudjema, A.~Pukhov and A.~Semenov, \emph{{micrOMEGAs 3: A
  program for calculating dark matter observables}},
  \href{https://doi.org/10.1016/j.cpc.2013.10.016}{\emph{Comput. Phys. Commun.}
  {\bfseries 185} (2014) 960}
  [\href{https://arxiv.org/abs/1305.0237}{{\ttfamily 1305.0237}}].

\bibitem{planck18}
{\scshape Planck} collaboration, N.~Aghanim et~al., \emph{{Planck 2018 results.
  VI. Cosmological parameters}},
  \href{https://arxiv.org/abs/1807.06209}{{\ttfamily 1807.06209}}.

\bibitem{LTHahn1998}
T.~Hahn and M.~Perez-Victoria, \emph{{Automatized one loop calculations in
  four-dimensions and D-dimensions}},
  \href{https://doi.org/10.1016/S0010-4655(98)00173-8}{\emph{Comput. Phys.
  Commun.} {\bfseries 118} (1999) 153}
  [\href{https://arxiv.org/abs/hep-ph/9807565}{{\ttfamily hep-ph/9807565}}].

\end{thebibliography}\endgroup

\end{document}